\documentclass[a4paper,11pt]{article}
\pdfoutput=1

\usepackage{jheppub} 

\usepackage[utf8]{inputenc}
\usepackage{placeins}
\usepackage{lineno}
\usepackage{amsmath}
\usepackage{amsfonts}
\usepackage[colorlinks=true,urlcolor=blue,citecolor=blue]{hyperref}
\usepackage{graphicx}

\begin{document}

\newcommand{\ie}{{\sl i.e.}~}
\newcommand{\eg}{{\sl e.g.}~}

\title{Unfolding measurement distributions via quantum annealing}

\author[1]{Kyle Cormier}
\author[1]{Riccardo Di Sipio}
\author[1,2,3,4,5]{Peter Wittek}

\affiliation[1]{University of Toronto, M5S 3E6 Toronto, Canada}
\affiliation[2]{University of Toronto, M5S 3E6 Toronto, Canada}
\affiliation[3]{Creative Destruction Lab, M5S 3E6 Toronto, Canada}
\affiliation[4]{Vector Institute for Artificial Intelligence, M5G 1M1 Toronto, Canada}
\affiliation[5]{Perimeter Institute for Theoretical Physics, N2L 2Y5 Waterloo, Canada}

\abstract{
High-energy physics is replete with hard computational problems and it is one of the areas where quantum computing could be used to speed up calculations. We present an implementation of likelihood-based regularized unfolding on a quantum computer. The inverse problem is recast in terms of quadratic unconstrained binary optimization (QUBO), which has the same form of the Ising hamiltonian and hence it is solvable on a programmable quantum annealer. We tested the method using a model that captures the essence of the problem, and compared the results with a baseline method commonly used in precision measurements at the Large Hadron Collider (LHC) at CERN. The unfolded distribution is in very good agreement with the original one. We also show how the method can be extended to include the effect of nuisance parameters representing sources of systematic uncertainties affecting the measurement.
}

\maketitle

\section{Introduction}
Problems in experimental high-energy physics have always challenged information and communication technologies.
The Large Hadron Collider at CERN is a great example of this, and with its upgrade program it is crucial that the computational resources continue scaling~\cite{dimeglio2017cern}.
Quantum computing is one the technologies that holds the promise of speeding up computationally expensive tasks.\footnote{Quantum Computing for High Energy Physics Workshop~\url{https://indico.cern.ch/event/719844/}}
Progress has been rapid in the development of quantum computing hardware, but implementations remain imperfect. Noise and limited scalability are the primary issues, leading to the term noisy, intermediate-scale quantum (NISQ) era~\cite{preskill2018quantum}.
Given the potential of quantum computing, there has been some exploratory work on using quantum devices in high-energy physics, for instance, in finding the Higgs boson in a simplified search domain~\cite{mott2017solving}, reconstruction of the trajectory of charged particles \cite{Bapst:2019llh} and calculation of quantum properties of final state radiation \cite{Bauer:2019qxa}.
This proof-of-concept was shown on a programmable quantum annealer~\cite{Johnson2011Quantum,tanaka2017quantum}, which is the largest scale quantum computer currently available.
We use the same architecture to apply quantum computing in a practical problem and solve a frequently appearing task in experiments: unfolding.

\begin{figure}[hbtp]
    \centering
    \includegraphics[width=\linewidth]{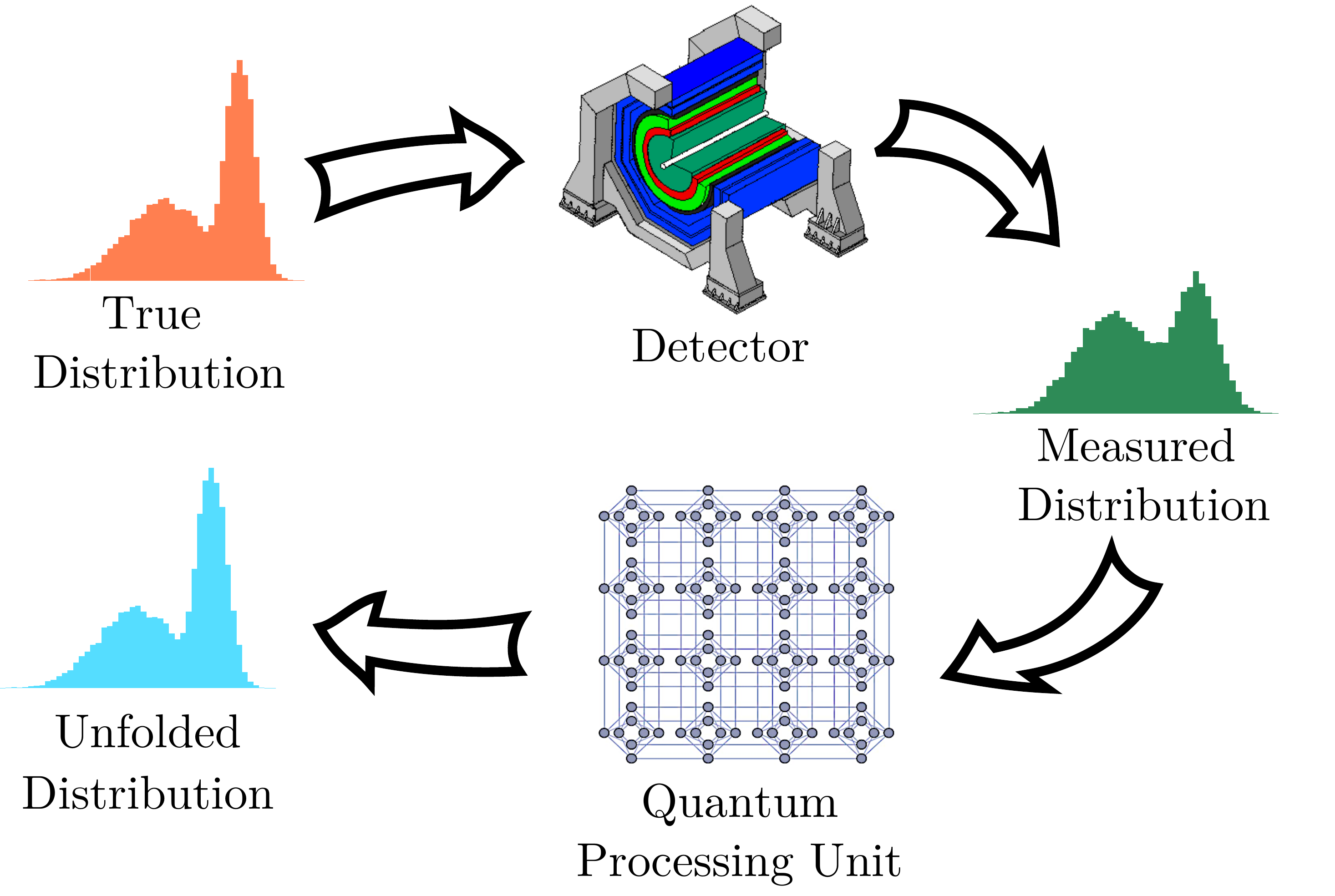}
    \caption{\small{The unfolding procedure corrects measured distributions, removing distortions caused by the detector. Here, we implement unfolding on a quantum computer~\cite{detectorImage}.}}
    \label{fig:overview}
\end{figure}

Unfolding is the procedure of correcting for distortions due to limited resolution of the measuring device~\cite{cowan1998statistical} as found in particle physics~\cite{aaboud2018measurements,sirunyan2019measurement}, astronomy~\cite{hogbom1974aperture} and other fields. This mathematical treatments is also known as the {\sl inverse problem} or {\sl deconvolution}. Unfolding is not necessary if the only goal is to compare the theory with the experimental results. In this case, a simulation of the experimental apparatus is used to account \eg for the interaction of radiation with matter, nuclear interactions, lens distortions, etc. In practical terms, this is usually carried out using computer codes such as Geant4~\cite{agostinelli2003geant}, FLUKA~\cite{battistoni2011fluka}, and Delphes~\cite{favereau2014delphes3}.
On the other hand, unfolding is essential if the aim is to compare measurements coming from different experiments. In general, each experimental apparatus has a unique signature in terms of detection efficiency, geometric acceptance and resolution. The overall effect is that the numerical value of some quantity generated in a given physical process is changed by the process of measurement. Unfolding is the mathematical procedure to ``correct'' for these effects and recover the original value.
Solving unfolding is a computationally challenging task: in what follows, we demonstrate how to map the problem to a quantum computer to accelerate the computations with upcoming quantum hardware. Fig.~\ref{fig:overview} outlines the proposed scheme.

Usually, the physical observable $\theta$ being measured is distributed with a probability density function $f(\theta)$ whose functional form is not known {\sl a priori}. It is customary to make use of binned distributions in the form of normalized histograms. Then, probability $p_j$ of finding $\theta$ in the bin $j$ is simply given by the integral of $f(\theta)$ in that bin, \ie:

\begin{equation}
    p_j = \int f(\theta)d\theta.
\end{equation}

The sum of all $p_j$ is equal to one. This probability distribution is turned into an actual physical prediction by multiplying the value of each bin by a dimensionful quantity such as a cross-section. Typically, this probability density function is estimated by simulating the physics process multiple times, using Monte Carlo methods, as it is done for example in so-called event generators such as Pythia~\cite{sjostrand2008pythia} or Herwig~\cite{bellm2016herwig}. The result of this operation is a histogram representing the ``true'' distribution $\theta$. After applying the effect of the experimental apparatus, one obtains another histogram corresponding to the ``reconstructed'' distribution $\mu$, \ie a prediction that can be compared to a measured distribution of real data $d$.

In its simplest form, the method to unfold the measurement to the truth level is to estimate efficiency factors $\epsilon_j$ for each bin from the simulation:

\begin{equation}
  \epsilon_j =  \mu_j / \theta_j.
\end{equation}

The unfolded distribution $\hat{\theta}$ is thus obtained by applying these efficiency factors to the observed data $d$ such that:

\begin{equation}
    \hat{\theta}_j = \epsilon^{-1}_j d_j.
\end{equation}

This method, called {\sl bin-to-bin correction}, has strong limitations: the histograms representing distributions before and after interacting with the experimental apparatus (in the following referred to as truth- and reco-level respectively) must have the same binning, correlations between adjacent bins are neglected, and there is no way to account for migration of events generated in a truth bin $j$ but ending up in a reco bin $i$.

These difficulties can be overcome if we promote the efficiency correction factors $\epsilon_j$ to a matrix $R_{ij}$ (usually called {\sl response matrix}), estimated from the simulation, such that:
\begin{eqnarray}
    \mu & = & R\theta \\
    \hat{\theta} &=& R^{-1}d. \label{eq:unreg_unfolding}
\end{eqnarray}

In principle, the number of bins at reco and truth level do not have to match, allowing also for over- and under-constrained unfolding.\footnote{Generally, such an equation does not have a unique solution if there are fewer (under-constrained) or more (over-constrained) parameters to fit at truth-level than bins at reco-level.} Typically, $R$ is presented as a matrix of (reco, truth) with rows that are normalized to unity, so that each entry is the probability that events produced in a given truth bin $j$ will be observed in the reco bin $i$. The inverse of this matrix multiplication, as applied in Eq.~\ref{eq:unreg_unfolding}, is commonly known as {\sl unregularized matrix inversion unfolding}.

Despite its relative simplicity, this method still suffers from numerical instabilities due to the inversion of the response matrix. This can be exacerbated by limitations in the simulations that may require very large computational resources to estimate off-diagonal elements with a high degree of precision. These elements encode very unlikely situations where the limited  resolution causes migrations, or a dramatic drop of efficiency or acceptance due to incomplete hermiticity, lack of instrumentation and other effects.
One possible approach is to constrain the truth-level distribution $\hat{\theta}$ by maximizing a likelihood function $\mathcal{L}(\mu|d)$ that depends on the estimated reco-level spectrum $\mu = \mu(\hat{\theta})$ and observed data $d$. In the case of a counting experiment, the likelihood is usually the product of Poisson or Gaussian (in the limit of large $\mu$) distributions for each bin.
Ideally, the truth-level unfolded distribution $\hat{\theta}$ is expected to show a good degree of regularity. For example, there should not be sharp transitions unless otherwise anticipated from a theoretical model. Among various possibilities, the most common way to achieve this is to impose an additional constraint in the form of a Lagrange multiplier added to the likelihood function, whose effect is to favour smooth solutions. The strength of the regularization can be controlled by an additional parameter $\beta$. A common measure of smoothness to be minimized is the second derivative of the distribution. This is an example of what is known more generally as {\sl Tikhonov regularization}. A similar approach to regularization is employed in the {\sl TUnfold} algorithm \cite{Schmitt_2012}. 

The complete likelihood to be maximized is thus:
\begin{eqnarray}
\label{eq:lhood}
\mathcal{L}(\mu|d) & = & \left ( \prod_i^N Poiss(\mu_i, d_i) \right ) \times e^{- \beta \rho}, \\ 
Poiss(\mu_i, d_i) & = & \frac{(\mu_i)^{d_i}}{d_i!}e^{-\mu_i},
\\
\mu_i & = & \sum_{j}^N R_{ij}\hat{\theta}_j, \\
\rho & = & \sum_{j=2}^{N-1} \left ( \hat{\theta}_{j+1} + \hat{\theta}_{j-1} \right )^2.
\end{eqnarray}

For numerical stability reasons, it is often preferable to minimize the logarithm of the likelihood, $\log \mathcal{L}(\mu|d)$. More advanced methods~\cite{dagostini1995multidimensional,choudalakis2012fully} have been developed to integrate out {\sl nuisance parameters}, \ie parameters needed to account for extra variation in the model such as sources of systematic uncertainty. 

\section{QUBO formulation of the unfolding problem}
To translate matrix operations to the quantum computing realm, an intermediate step has to be taken: the likelihood function presented in Eq. \ref{eq:lhood} has to be adapted so that the unfolding can be expressed in terms of a binary optimization, which in turn can be implemented on a quantum annealer.

The quadratic unconstrained binary optimization (QUBO) model requires the minimization of an objective function:
\begin{equation}
    y = q^T C q,
\end{equation}
where $q$ is a vector of binary elements (observing that $q_j^2 = q_j$ if $q_j$ can only be either 0 or 1) and $C$ is a  square matrix of constants, often presented in upper triangular form. QUBO is a special case of quadratic programming. Usually, the minimization is subject to a number of constraints, which can be encoded in a quadratic penalty term, such that:

\begin{eqnarray}
y & = & q^T C q =\nonumber \\
  & = & \sum_{j=1}^N c_{jj} q_jq_j + \sum_{1\leq j < k \leq N} c_{jk}q_j q_k.
  \label{eq:qubo}
\end{eqnarray}

The linear constraints appear on the diagonal of the $C$ matrix, while the off-diagonal elements encode the quadratic penalty terms.

To rewrite the unfolding equation, we follow the approach described in Ref.~\cite{omalley2016}. First, we replace the Poisson term with a sum of squares of differences, \ie the square of the $L_2$ norm between the observed data and the reconstructed spectrum, which corresponds to taking the Gaussian approximation. Then, the regularization is obtained by minimizing the second derivative up to a multiplicative factor $\lambda$, which is obtained by multiplying the truth-level spectrum $x$ by the Laplacian operator $D$. Thus, the objective function to be minimized is:

\begin{equation}
    y = \left \| Rx - d \right \|^2 + \lambda  \left \| Dx \right \|^2.
    \label{eq:objective1:matrixform}
\end{equation}

The detailed conversion of Eq. \ref{eq:objective1:matrixform} into QUBO form can be found in the appendix. 
Among some other algebraic manipulations, we have to deal with the fact that the weights above have been derived for a real-valued vector $x$ (of $N$ elements) rather than a binary-value vector $q$ of $nN$ elements, where $n$ is the number of bits (typically a power of 2) used to encode the elements of the real-valued vector. This can easily resolved by constructing equivalent matrices that act upon the discretized version of $x$. This can be achieved by expanding $x$ as:

\begin{equation}
    x_i = \alpha_{i} + \beta_{i}\sum_{j=0}^{n-1} 2^{j}q_{n\times i + j}.\label{eq:encoding1}
\end{equation}

The QUBO is formulated from Equation~\ref{eq:encoding1}, by replacing $x_i$ with its binary encoding.

This way, the QUBO minimization can be expressed in terms of the binary vector $q$:

\begin{equation}
    y = \left \| R_2q - d \right \|^2 + \lambda  \left \| D_2q \right \|^2.
\end{equation}

\section{Systematic uncertainties}
Sources of systematic uncertainty shift the value of each bin by a certain amount. Typically, variations corresponding to $\pm$1 standard deviation are estimated from the simulation, and an interpolation is done during the optimization. For each bin $i$, the effect of the $k$-th systematic is represented by an additional term, so that the predicted spectrum is:

\begin{equation}
    y^\prime_i = y_i + \Delta y_{i} =
     R_{ij}x_j + \sum_k^{N_{syst}} z_k s_{ik}.
\end{equation}

where $y$ is the central prediction, $s_{ik}$ is the variation in bin $i$ due to systematic $k$ of strength $z_k$.

To take the systematic effects into account, an extended vector ($\tilde{x}$) which includes the values of the systematic strengths is defined. Equally, we define extended matrices $\tilde{R}$ and $\tilde{D}$:

\begin{eqnarray}
   \tilde{x} & = & ( x_1,\hdots, x_N\ \vline\ z_1, \hdots, z_K ),\\
   \tilde{R} & = &
   \begin{pmatrix}
   R_{11} & \hdots & R_{1N} & \vline & s_{11} & \hdots & s_{1K} \\
   \vdots &  \ddots & \vdots & \vline & \vdots & \ddots & \vdots \\
   R_{N1} & \hdots & R_{NN} & \vline & s_{N1} & \hdots & s_{NK}
   \end{pmatrix},\nonumber \\
      \tilde{D} &=& 
   \begin{pmatrix}
   D_{11} & \hdots & D_{1N} & \vline & 0 & \hdots & 0 \\
   \vdots &  \ddots & \vdots & \vline & \vdots & \ddots & \vdots \\
   D_{N1} & \hdots & D_{NN} & \vline & 0 & \hdots & 0
   \end{pmatrix}.\nonumber
\end{eqnarray}

By replacing $x$, $R$, and $D$ in Equation~\ref{eq:objective1:matrixform} with $\tilde{x}$, $\tilde{R}$, and $\tilde{D}$, the values of the systematic uncertainties can be altered and the predicted spectrum will change accordingly. Generally, such an equation does not have a unique solution as it corresponds to an under-constrained problem (there are more parameters to fit at truth-level than bins at reco-level). 

An additional matrix, $S$, defined by: 

\begin{equation}
S \tilde{x} = 
\gamma
\begin{pmatrix}
\begin{matrix}
0 &         & \\
   & \ddots & \\
   &        & 0
\end{matrix}
& \vline &
\begin{matrix}
1 &         & \\
   & \ddots & \\
   &        & 1
\end{matrix}
\end{pmatrix}
\begin{pmatrix}
   x_1 \\ \vdots \\ x_N \\ \hline z_1 \\ \vdots \\ z_K
   \end{pmatrix}.
\end{equation}

is introduced to penalize large systematics. The size of the penalty is controlled by an adjustable parameter $\gamma$. The final objective function to be minimized is then given by:

\begin{equation}
\label{eq:obj:syst}
    y = \left \| \tilde{R}\tilde{x} - d \right \|^2 + \lambda \left \| \tilde{D}\tilde{x} \right \|^2 + \gamma \left \| S \tilde{x}\right \|^2 .
\end{equation}

In this case, the systematic shifts $z_k$ are being minimized simultaneously to the spectrum, with a penalty imposed for any deviation from their nominal value of 0, with a strength controlled by the parameter $\gamma$. The form of the QUBO weights which include these additional terms is given in the appendix.

\section{Quantum Annealing}
Many NISQ-era quantum devices such as programmable quantum annealers~\cite{Johnson2011Quantum}, variational quantum eigensolvers~\cite{Peruzzo2014} and variational circuits~\cite{farhi2014qaoa} on gate-model quantum computers, or networks of degenerate optical parametric oscillators~\cite{qnncloud2017qnn} are proposed for solving QUBO tasks~\cite{denchev2016what,Crosson2016,mandra2018deceptive}.
In most cases, the QUBO is mapped to a classical Ising model with a simple change of variables.
This includes the D-Wave 2000Q quantum annealer with 2038 manufactured spins that our experiments relied on.
On this device, the coupling and onsite fields of the Ising model are bounded the following way:

\begin{eqnarray}
  \hat{H} =  \sum_{i,j} J_{ij}\hat{\sigma}^z_i \hat{\sigma}^z_j + \sum_i h_i \hat{\sigma}^z_i\\
J_{ij} \in  [-1,1],  h_i  \in  [-2,2] \nonumber,
\end{eqnarray}
where $\hat{H}$ is the cost Hamiltonian for which the annealer attempts to find the ground state.

\section{Results}
We tested the method described above in a realistic situation, where the quantity to be unfolded is represented by a histogram with 5 bins and migrations are not negligible (\ie{} non-zero off-diagonal elements in the response matrix).\footnote{The code is available under an open source license in the code repository \url{https://github.com/rdisipio/quantum_unfolding}} We assume that the distortions introduced by the detector are perfectly known. Hence, in order to obtain the corresponding reco-level histograms, we applied the same response matrix to the truth-level distributions of both $\mu$ and $d$, which are different in the most general case. The pseudo-data reco-level histogram is then unfolded using the following methods: D'Agostini iterative Bayesian~\cite{dagostini1995multidimensional} with $N_{itr}=4$; QUBO solved on the CPU by the {\sl neal} simulator \cite{neal}; and QUBO executed by the real DWave QPU, with the regularization strength $\lambda$ set values between 0 and 1. The number of iterations in the benchmark unfolding method was set to 4 as a compromise between bias and statistical uncertainty. This setting is a common choice in measurements at the LHC~\cite{aaboud2018measurements,cms2017diffxs}. The uncertainty is calculated from the square root of the diagonals of the covariance matrix given by the unfolding, as implemented in the RooUnfold package~\cite{RooUnfold}. The uncertainty associated to the unfolding quantum annealing represents one standard deviation calculated on a set of 20 executions with 5,000 reads from the QPU per run.

\begin{figure}[hbtp]
    \centering
    \includegraphics[width=\linewidth]{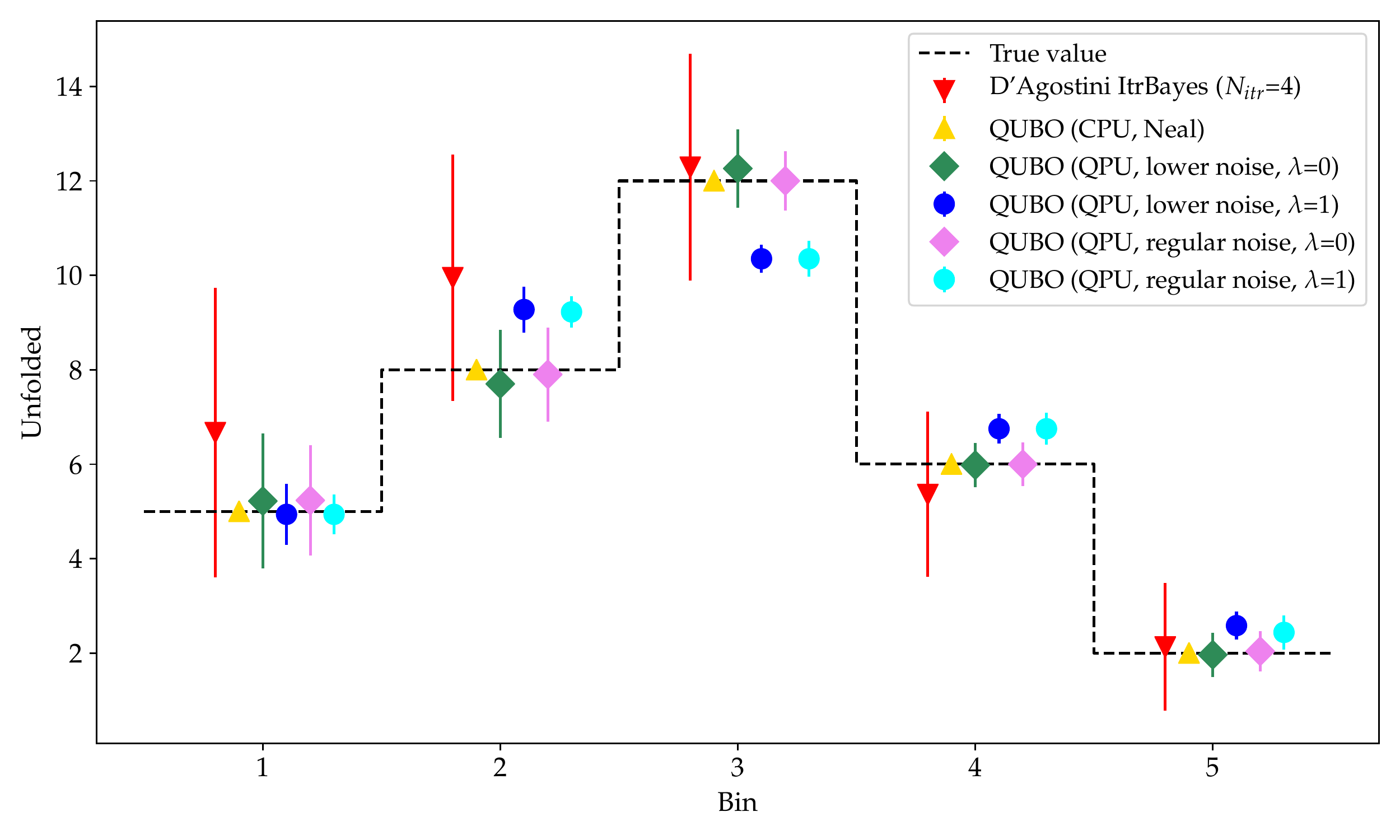}
    \caption{\small{Truth-level distribution of pseudo-data, unfolded with different methods, of a peaked spectrum. The data corresponding to runs submitted to the QPU are averaged over 20 executions with 5000 reads each.}}
    \label{fig:unfolded:peak}
\end{figure}

A first test is shown in Fig. \ref{fig:unfolded:peak}, where a peaking distribution is unfolded. This is representative of the mass spectrum in presence of a resonance. All the unfolding methods are able to recover the truth-level distribution, with an agreement always in the order of about one standard deviation. There is no apparent difference between the lower-noise annealer DW\_2000Q\_5 solver and the older regular-noise DW\_2000Q\_2\_1. 

\begin{figure}[hbtp]
    \centering
    \includegraphics[width=\linewidth]{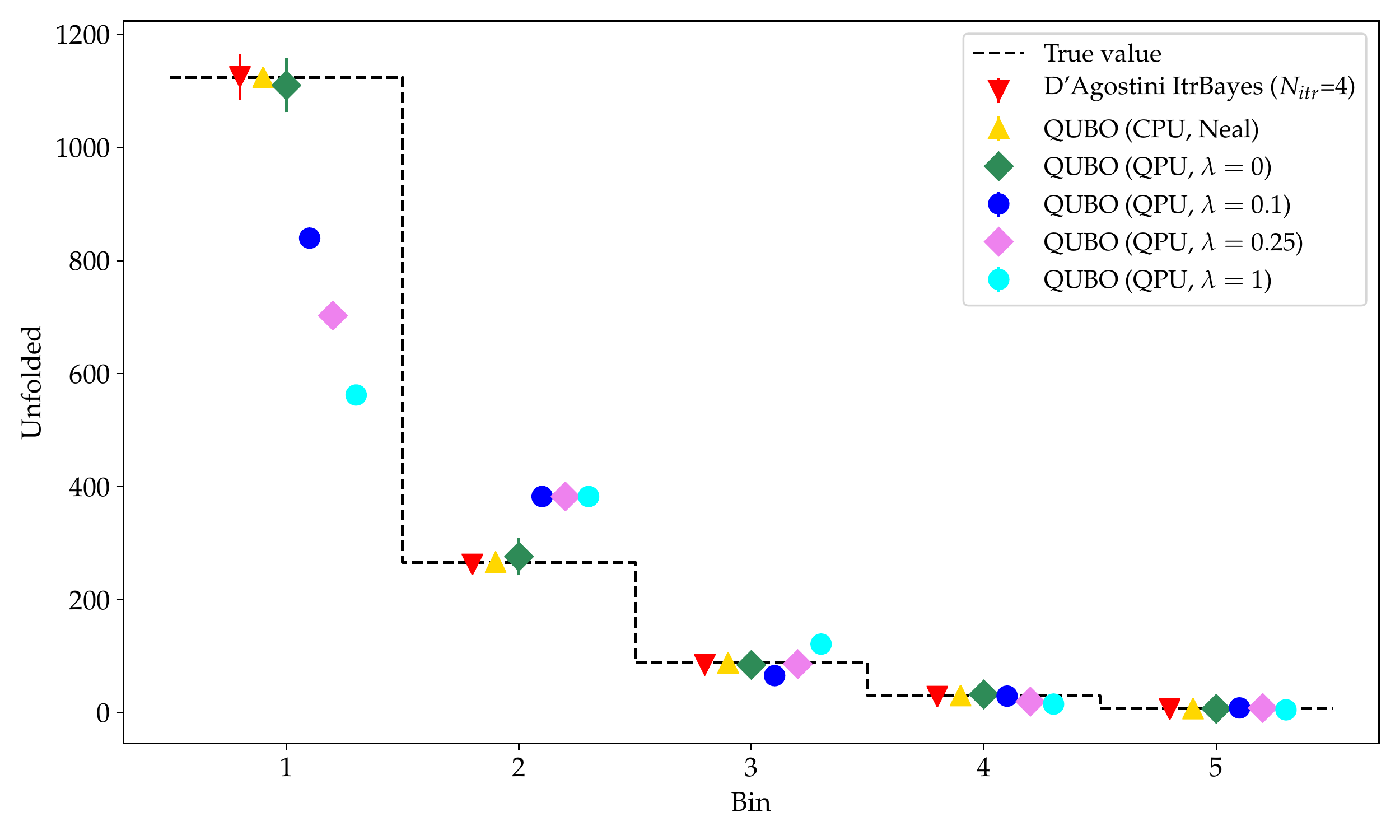}
    \caption{\small{Truth-level distribution of pseudo-data, unfolded with different methods and values of the regularization strength $\lambda$, of a steeply-falling spectrum. The data corresponding to runs submitted to the QPU are averaged over 20 executions with 5000 reads each.}}
    \label{fig:unfolded:falling}
\end{figure}

Fig. \ref{fig:unfolded:falling} shows the unfolding applied to a steeply falling spectrum, representative of a certain class of observables such as the transverse momentum of a particle. In this case, the unregularized QUBO is in good agreement with both the truth distribution and the D'Agostini method within one standard deviation. The impact of the regularization is evident as larger and larger values of the $\lambda$ parameter result in flatter and flatter solutions which do not agree well with the truth level. This is a confirmation of the well-known fact that the regularization is problem-dependent, and an optimal value of its strength has to be carefully estimated \eg~from the simulation.

The good agreement with the simulated annealing and with the D'Agostini method indicates that with a large enough numerical precision, the QUBO unfolding would have a comparable performance with the traditional methods. On the other hand, the execution on the QPU is less accurate due to hardware limitations. 

\begin{figure}[hbtp]
    \centering
    \includegraphics[width=\linewidth]{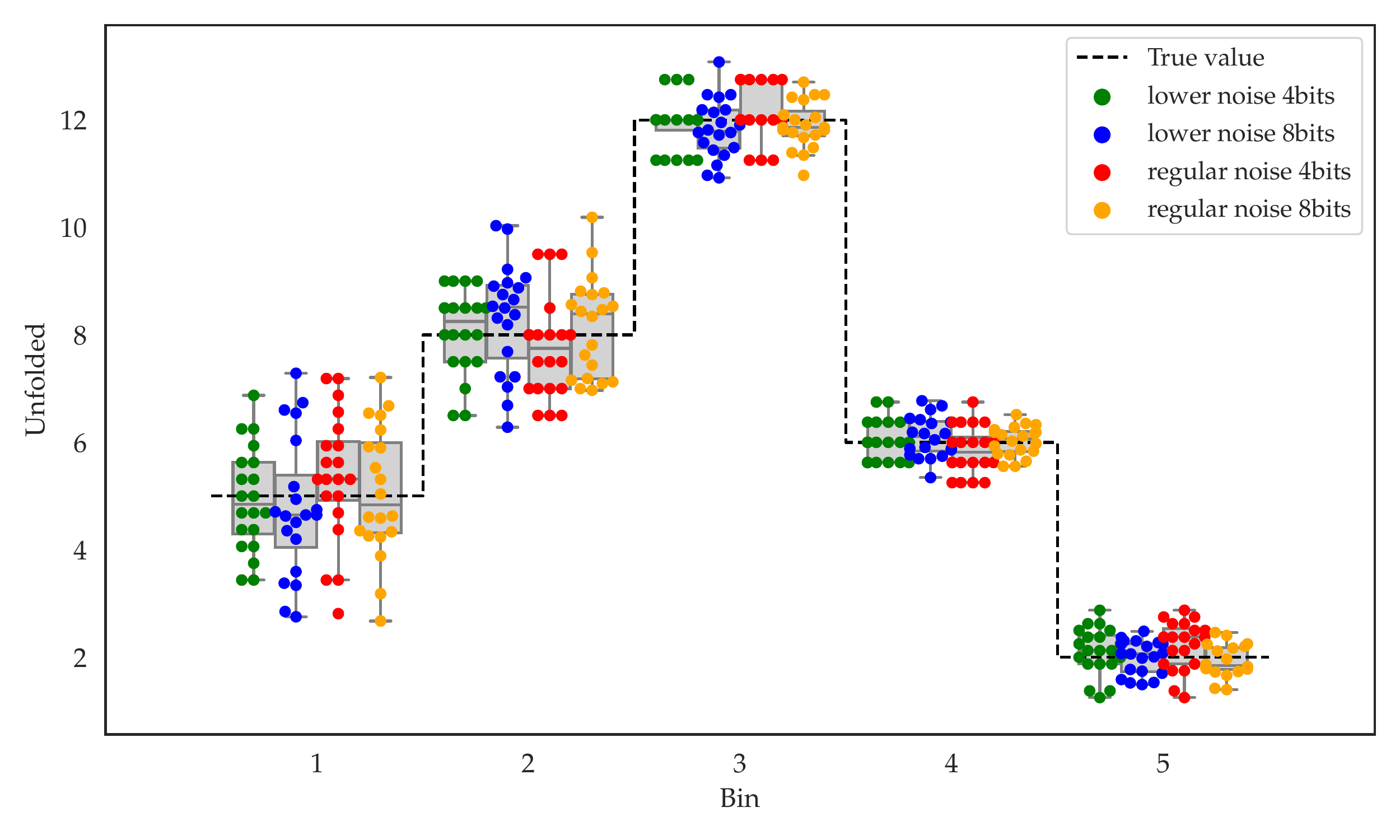}
    \caption{\small{Truth-level distribution of pseudo-data, unfolded with either a 4- or 8-bits encoding, of a peaked spectrum. The data corresponding to runs submitted to the QPU are averaged over 20 executions with 5000 reads each. The maximum chain length in the best embedding is 7 and 15 qbits respectively. The 8-bit encoding allows a more fine-grained estimation of the unfolded histogram.}}
    \label{fig:unfolded:falling:nbits}
\end{figure}

A further test was performed to compare the difference between a 4-bit and 8-bit encoding. The result is shown in Fig.~\ref{fig:unfolded:falling:nbits}. In principle, with more bits it should be possible to have a better numerical precision. However, it also results in longer chains of qbits in the minor embedding.  

\begin{figure}[hbtp]
    \centering
    \includegraphics[width=\linewidth]{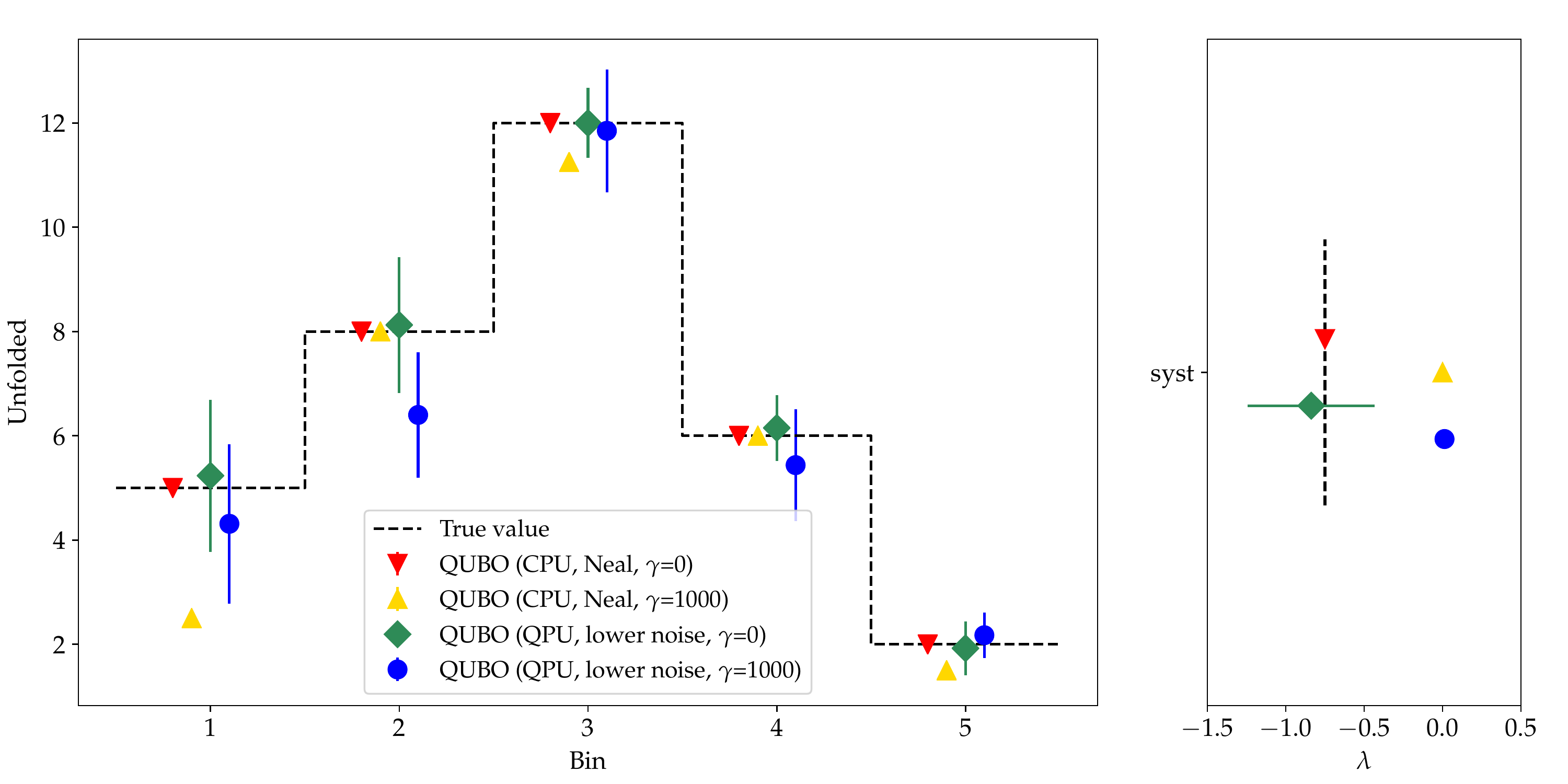}
    \caption{\small{Truth-level distribution of pseudo-data, unfolded with different methods, of a peaked spectrum including the effect of a shape systematic. The data corresponding to runs submitted to the QPU are averaged over 20 executions with 5000 reads each.}}
    \label{fig:unfolded:peak:syst}
\end{figure}

To test the effect of systematic uncertainties, we added one component to the pseudo-data: a shape systematic whose effect is to distort the number of entries in each bin. We set the strength of this systematic to -0.75. The unfolding is shown in Fig. \ref{fig:unfolded:peak:syst}. The discretized nature of the optimization introduces some  interesting features. For this particular setup, the full solution is not degenerate, even when no penalty is applied for increasing the systematic uncertainty (i.e. $\gamma = 0$). In this case, the simulated annealing correctly picks the value of the systematic and each bin of the distribution. The QPU also picks out this solution on average. 
 
If a sufficiently large penalty parameter is applied (in this case, $\gamma$=1000), the annealing prefers a value of the systematic strength close to zero. In order to accommodate this choice, the bins gets shifted. The solution found by the simulated annealer in this case matches the brute force solution, however this solution is not found by the QPU. While more work remains to understand many subtleties, a more refined version of this methodology could be used to estimate the impacts of systematic uncertainties. This is an area for continued work.

Finally, we used the DWave-Hybrid software framework to find the solution by  searching for the state of minimum energy using a Tabu heuristic algorithm~\cite{glover1989tabu,glover1990tabu} and the QPU in parallel. The result we obtained is completely in agreement with both the truth-level distribution and the one obtained with the simulated annealing. 
While the hybrid approach goes well beyond the needs of the scenario under consideration, we argue that this is likely to be the key for a typical measurement at the LHC where the number of bins is larger than 10 and the number of sources of systematic uncertainty is on the order of 100. 

\section{Conclusions}
NISQ-era quantum devices are meant to work in close integration with classical computing resources, creating hybrid classical-quantum algorithms.
We found that using the quantum annealer together with a classical optimization metaheuristics helps with two problems: (1) problems larger than the hardware spin system can be solved; (2) results are quickly refined into the optimal solutions in a few iterations between the quantum hardware and the classical algorithm.
The close agreement between the correct results and the one obtained by the hybrid classical-quantum protocol indicates the relevance of this paradigm to practical problems in experimental high-energy physics.

Upcoming quantum hardware, like the Pegasus architecture that has a denser connectivity and over five thousand qubits~\cite{dattani2019pegasus}, could easily tilt the balance towards the quantum part of the protocol, yielding faster time to solution compared to a purely classical solver.
The scheme we introduced here may also be amenable to taking into account discretization effects, which can be significant in problems with low statistics~\cite{gaponenko2019practical}. The test case presented in this paper was limited to five bins in order to fit in the currently available hardware. However, the above considerations indicate that it will be possible to unfold histograms with more bins,  a higher precision due to an encoding with more bits, and a larger number of systematics by the means of the integration of more powerful and less noisy QPUs with classical systems based on CPUs. 

\section*{Acknowledgments}
We acknowledge useful discussions with H. Sadeghi and A. Condello (D-Wave Systems).
This research was supported by Perimeter Institute for Theoretical Physics. Research at Perimeter Institute is supported by the Government of Canada through Industry Canada and by the Province of Ontario through the Ministry of Economic Development and Innovation. We also acknowledge the support of the Natural Sciences and Engineering Research Council of Canada (NSERC).

\FloatBarrier

\bibliographystyle{ieeetr}
\bibliography{main}

\section*{Appendix}

This appendix gives the detailed derivation of mapping the unfolding problem to a quadratic unconstrained binary optimization (QUBO). After deriving the basic formulation with Tikhonov regularization, we add an additional regularization term for systematic uncertainties.

\emph{Deriving the QUBO formulation of the unfolding problem}
To rewrite the unfolding equation, we follow the approach described in Ref.~\cite{omalley2016}. First, we replace the Poisson term with a sum of squares of differences, \ie the square of the $L_2$ norm between the observed data and the reconstructed spectrum. Then, the regularization is obtained by minimizing the second derivative, which is obtained by multiplying the truth-level spectrum $x$ by the Laplacian operator, \eg:

\begin{equation}
D =
\begin{pmatrix}
  -2 & 1  & 0 & 0 & 0\\
  1  & -2 & 1 & 0 & 0\\
  0  & 1  & -2 & 1 & 0\\
  0 & 0 & 1 & -2 & 1 \\
  0 & 0 & 0 & 1 & -2
\end{pmatrix}.
\end{equation}

Other regularization choices are possible, but the second derivative is a common choice, and used for the calculations in this paper. The relative strengths of the regularization term (as compared to the agreement between the data and the reconstructed spectrum) is controlled by a parameter, $\lambda$. Thus, the objective function to be minimized is:

\begin{equation}
    y = \left \| Rx - d \right \|^2 + \lambda  \left \| Dx \right \|^2.
    \label{eq:objective:matrixform}
\end{equation}

To convert Eq. \ref{eq:objective:matrixform} into QUBO form such as Eq. (12) in the main text, we rewrite the equations using index notation, i.e.:

\begin{eqnarray}
 Ax & \to & A_{ij}x_j, \\
 AB & \to & A_{ij}B_{jk}, \\
 AB^T & \to & A_{ij}B_{ik}. 
\end{eqnarray}

A summation over repeated indices is implicit, unless otherwise noted. Thus, we have:

\begin{eqnarray}
 \label{eq:obj:2}
 y & = & \left ( R_{ij}x_j - d_i \right )\left ( R_{ik}x_k - d_i \right ) + \lambda \left ( D_{ij}x_j D_{ik}x_k \right )  \nonumber\\
   & = & \left ( R_{ij}R_{ik} + \lambda D_{ij}D_{ik} \right )x_j x_k - 2 R_{ij}x_j d_i + d_i d_i  \label{eq:obj:1} \nonumber\\
\end{eqnarray}

The term $d_i d_i$ can be ignored, since it adds a constant, which does not affect the optimization. Before determining the QUBO weights, we need to encode the floating point values $x_i$ into an appropriate binary format for the QUBO.

The simplest encoding, which is suitable for some situations, is a standard n-bit integer encoding:

\begin{equation}
    x_i = \sum_{j=0}^{n-1} 2^{j}q_{n\times i+j},
\end{equation}

where $q_a \in \{ 0, 1\}$.
However, this is somewhat limiting for two reasons: firstly, this limits the values to positive integers; secondly, current quantum computing hardware requires the bit encodings to be small, which may put limitations on the range of values which can be encoded.

Standard floating point encodings do not lend themselves naturally to the QUBO problem because the exponent bits give rise to non-quadratic terms.
Here, we chose to use a non-standard encoding. Each value $x_i$ is encoded  with $n$ bits using an offset, $\alpha_{i}$, and a scaling parameter, $\beta_{i}$:

\begin{equation}
\label{eq:encoding}
    x_i = \alpha_{i} + \beta_{i}\sum_{j=0}^{n-1} 2^{j}q_{n\times i + j}.
\end{equation}

The values of the offset and scaling parameters are stored on the classical computer, so that they can be used to encode and decode the the QUBO results. 
In order to provide additional flexibility, each value in $\vec{x}$ may be encoded with a different number of bits. This flexibility is implemented in the code, but for clarity of exposition, the fixed-bit encoding is given here. To simplify the notation, we define $\beta_{ia}$ instead of writing the sum over $j$: 

\begin{equation}
    x_i = \alpha_{i} + \beta_{ia}q_{a}.
\end{equation}

Defining $a_0 \equiv i\times n$, we can reproduce Equation~\ref{eq:encoding} by setting $\beta_{ia}= \beta_{i}2^{(a-a_0)}$ for $ 0 \leq (a- a_0) < n$ and 0 otherwise.
The QUBO is formulated from Equation~\ref{eq:obj:2}, by replacing $x_i$ with its binary encoding. Additionally, for notational convenience, we define $W_{jk} \equiv (R_{ij}R_{ik} + \lambda D_{ij}D_{ik})$.
This gives:

\begin{eqnarray}
\label{eq:objectivebinaryencoding}
    y &=& W_{jk}\alpha_{j}\alpha_{k} +
    2W_{jk}\alpha_{j}\beta_{ka}q_{a} \nonumber \\ 
    & & + W_{jk}\beta_{ja}\beta_{kb}q_{a}q_{b} \nonumber \\
    & &- 2R_{ij}d_{i}\alpha_{j} - 2R_{ij}d_{i}\beta_{ja}q_{a} + d_i d_i.
\end{eqnarray}

As before, constant terms (those not containing $q_a$) have no impact on the optimization; therefore, the terms $W_{jk}\alpha_{j}\alpha_{k}$,
 $R_{ij}d_{i}\alpha_{j}$, and $d_id_i$ can be ignored.  Expressing in terms of the binary values $q_a$, we also have $q_a q_a = q_a$. So that for $a=b$ the term $W_{jk}\beta_{ja}\beta_{kb}q_{a}q_{b}$ can be rewritten as simply $W_{jk}\beta_{ja}\beta_{ka}q_{a}$. Enforcing the requirement $a < b$ then Equation~\ref{eq:objectivebinaryencoding} can be rewritten in QUBO form as:
 
\begin{equation*}
y'  =  \sum_{a=1}^N c_{aa} q_aq_a + \sum_{ a < b } c_{ab}q_a q_b;
  \label{eq:qubo2} 
\end{equation*} 

with:

\begin{eqnarray}
 \label{eq:QUBO:quadratic_weights}
 c_{ab} &=& 2 W_{jk} \beta_{ja}\beta_{kb}, \\
 c_{aa} &=& 2 W_{jk} \alpha_{k}\beta_{ja} + \nonumber \\  
 & & W_{jk}\beta_{ja}\beta_{ka}  - 2R_{ij}d_i\beta_{ja}.
 \label{eq:QUBO:linear_weights}
\end{eqnarray}

The repeated index $a$ is not summed over in Equation~\ref{eq:QUBO:linear_weights}.

In general, determining the parameters $\alpha_i$ and $\beta_i$ is a problem-specific task. They define the space of possible solutions, and therefore can be chosen to force certain constraints. For example, for problems where $\vec{x} \in \mathbb{N}$ choosing $\alpha_{i}$, $\beta_i \in \mathbb{N}$ will ensure  the solution also respects this form. Nonetheless, the solution will only span a certain range. We use a heuristic to form initial guesses for suitable choices. Additional heuristics and criteria for updating the values based on results can be developed in a straightforward manner.

\emph{QUBO weights with systematic uncertainties}--- Only a minor modification of the above derivation is required to include the case of systematic uncertainties. The QUBO weights for the vector $\tilde{q}$, corresponding to the binary encoding of $\tilde{x}$ proceeds by replacing Equation~\ref{eq:obj:2} with Equation~\ref{eq:obj:syst} and following the  derivation above.  The result of the calculation is that the QUBO weights for $\tilde{q}$ are then as given in Equations~\ref{eq:QUBO:quadratic_weights} and~\ref{eq:QUBO:linear_weights} but with
$W_{ij} = \tilde{R}_{ij}\tilde{R}_{ij} + \lambda \tilde{D}_{ij}\tilde{D}_{ij} + \gamma S_{ij} S_{ij}$.

In fact,  Equation~\ref{eq:obj:syst} can be transformed into the form of Equation~\ref{eq:obj:2} by defining a matrix:

\begin{equation*}
\tilde{D'} = \begin{pmatrix}
 \tilde{D} \\
 \hline 
\sqrt{\frac{\gamma}{\lambda}}S
\end{pmatrix}
\end{equation*} 

which is of block diagonal form and treats the regularization and systematic uncertainties in a single term. Then, the form of $W_{ij}$ follows from direct substitution.

\end{document}